\documentclass[onecolumn,showpacs,preprintnumbers,amsmath,amssymb]{revtex4}
\usepackage[dvips]{graphicx}
\usepackage{bm}

\begin{document}
\title{Quasi-periodic oscillations in a network of four R\"{o}ssler chaotic
oscillators}
\author{Alexander~P.~Kuznetsov, Igor~R.~Sataev, Yuliya~V.~Sedova and Ludmila~V.~Turukina}
\affiliation{Kotel'nikov's Institute of Radio-Engineering and Electronics of RAS, Saratov Branch,\\
Zelenaya 38, Saratov, 410019, Russian Federation}

\date{\today}

\begin{abstract}
We consider a network of four non-identical chaotic R\"{o}ssler oscillators. The
possibility is shown of appearance of two-, three- and four-frequency
invariant tori resulting from secondary quasi-periodic Hopf bifurcations and
saddle-node homoclinic bifurcations of tori.
\end{abstract}

\pacs{05.45.Pq, 05.45.Xt}

\maketitle

\section{Introduction}
Problem of the interaction of oscillators of different nature is the focus
of researchers in different fields of physics, chemistry, biology. There are
examples from the domains such as optomechanical, micromechanical systems,
Josephson junctions, ion traps, radio oscillators, etc.~\cite{1,2,3,4,5,6,7,8}. Individual
oscillators can be periodic or chaotic. In the first case their weak
interaction leads to synchronization or quasi-periodic oscillations. If you
increase the number of oscillators in the system, then the number of
possible incommensurate frequencies is increasing too. The result may be
multi-frequency quasi-periodic oscillations, which correspond to invariant
tori of higher and higher dimension ~\cite{9,10,11,12,13,14,15}. An increase of coupling
parameter may lead to breakdown of such tori with the formation of chaos. In
this letter we discuss an alternative situation. We consider a network of a
small number of coupled chaotic oscillators, and demonstrate the occurrence
of invariant tori of different dimensions with increasing coupling
parameter.

The possibility of quasi-periodic oscillations in coupled chaotic system was
pointed out in ~\cite{16,17}. While investigating the ring of three
unidirectionally coupled identical Lorenz systems the modes were found of
not only two-, but also three-frequency quasi-periodicity. However, only
one-parameter analysis was carried out and the three-frequency tori were
fixed exceptionally in a very narrow range ($\sim $0.001) of the parameter
values. The emergence of a new frequency in ~\cite{16,17} was explained by the
possibility of rotational motion in the system and a high degree of symmetry
due to the identity of interacting subsystems .

Here we consider the case, which is more universal. As the basic model we
use a network of small number of globally coupled chaotic R\"{o}ssler
oscillators. The important point is the oscillators are not identical. It is
well known that two such oscillators demonstrate asynchronous chaotic
regimes and synchronized chaotic and periodic regimes ~\cite{18,19,20}. In ~\cite{19,20}
the "outline" of the picture of domains of different modes was presented on
the ``frequency detuning - coupling strength'' parameter plane. Increasing
the number of oscillators to four enriches and complicates this picture,
which requires more detailed study. Such a task is the subject of this
paper. A convenient tool for such studies is the Lyapunov analysis. Another
issue we will discuss here is the question of the possible type of
bifurcations of invariant tori. We have found quasiperiodic Hopf bifurcation
~\cite{21}, which corresponds to the soft emergence of higher dimension torus.
There is also possible the saddle-node homoclinic bifurcation, typical for a
resonant tori on the hypersurface of a torus of higher dimension.

\section{Results}
Let us consider a network of four coupled R\"{o}ssler oscillators
\begin{equation}
\label{eq1}
{\begin{array}{l}
 \dot {x}_n = - (1 + \frac{n - 1}{3}\Delta )y_n - z_n , \\
 \dot {y}_n = (1 + \frac{n - 1}{3}\Delta \mbox{)}x_n + py_n + \frac{\mu
}{3}\sum\limits_{i = 1}^4 {(y_i - y_n )} , \\
 \dot {z}_n = q + (x_n - r{ }_n)z_n . \\
 \end{array}}
\end{equation}
Here $\Delta $ parameter is introduced by analogy with ~\cite{9,10,11} and is
responsible for the frequency detuning of the oscillators. Values of the
parameters p = 0.15, q = 0.4, r=8.5 correspond to the chaotic regime in
individual subsystems ~\cite{18,19,20}.

The system (\ref{eq1}) may demonstrate hyperchaotic regimes with different numbers
of positive Lyapunov exponents, limit cycles of different types, as well as
the invariant tori of different dimensions. More detailed information about
the structure of $(\Delta ,\mu )$ parameter plane can be obtained by
constructing the Lyapunov exponents charts. In this case, at the each point
in the plane the Lyapunov spectrum is computed and then the plane is painted
in different colors in accordance with its signature ~\cite{12,13,14,15,21}.

Lyapunov exponents chart for the system (\ref{eq1}) is shown in Figure 1. Marked
with different colors are the periodic modes P, quasiperiodic T with
different number N of incommensurate frequencies, chaos C and hyperchaos CH
with different number of positive exponents. In addition, there is the
"amplitude death" mode AD, which is responsible for disappearance of
oscillations due to their suppression by dissipative coupling. The color
legend is under the figure. Numbers denote periods in the Poincare
section for the simplest limit cycles.
\begin{figure}[h!]
\includegraphics[height=12cm, keepaspectratio]{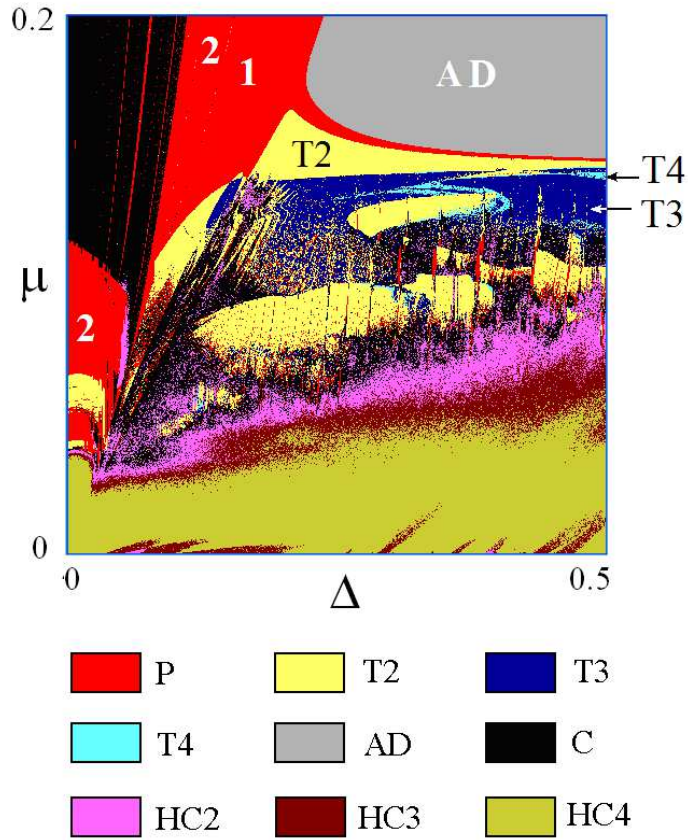}
\caption{Lyapunov exponents chart for the system (\ref{eq1}), $p$=0.15, $q$=0.4, $r =
8.5$.}
\end{figure}

We first discuss the region of non-synchronous modes, which corresponds to
sufficiently large frequency detuning. At low coupling the hyperchaotic mode
HC4 dominates with four positive Lyapunov exponents, which is quite natural
for a system of four coupled chaotic oscillators. With the increasing of
coupling the number of positive exponents gradually decreases, and after the
transition via HC2 and HC3 regims the usual chaos C arises. For even more
stronger coupling the complicated picture of alternating of modes of
different types may be observed, and typical are two-frequency T2 and
three-frequency T3 tori.

Examples of invariant tori in the Poincare section at $\Delta $ = 0.25 for
increasing coupling parameter in the $0.098 < \mu < 0.15$ range are given in
Figure 2. Poincare section was defined by conditions $y_1 = 0,\,\,x_1 > 0$.
It can be seen that for sufficiently strong coupling a simple torus may be
observed, in the Poincare section we see invariant curve close to a circle,
Figure 2a. With a decrease of coupling this curve is distorted in shape.
Then, three-frequency torus arises softly, Figure 2b. Next there is a
sufficiently large number of various regimes transformations. For example,
it can be seen on Figure 2c that the phase trajectories are condensating
along some directions inside the hypersurface of a three-frequency torus.
From these condensations the tangled two frequency resonant torus arises as
a result of the saddle-node homoclinic bifurcation, Fig. 2d. Note that the
transition from Fig.2c to Fig. 2d corresponds to a very small change in the
parameter. With a further reduction of coupling another saddle-node
homoclinic bifurcation of the resonant two-torus occurs, and there is again
a three-frequency torus. Number of resonant windows of this type is
sufficiently large.
\begin{figure}[h!]
\includegraphics[height=10cm, keepaspectratio]{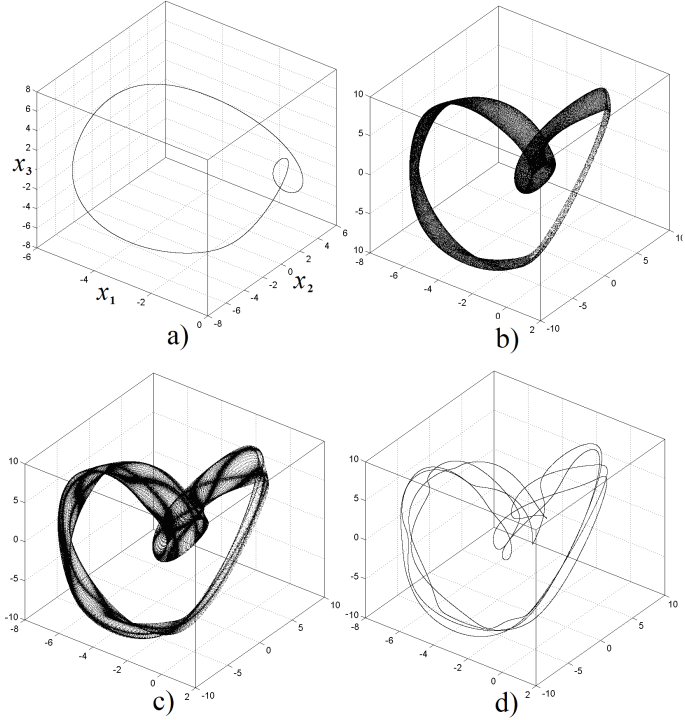}
\caption{Examples of Poincare sections of two- and three-frequency
invariant tori, $\Delta $=0.25; a) $\mu $=0.143, b) $\mu $=0.14, c) $\mu
$=0.13889, d) $\mu $=0.138919.}
\end{figure}
\begin{figure}[h!]
\includegraphics[height=6cm, keepaspectratio]{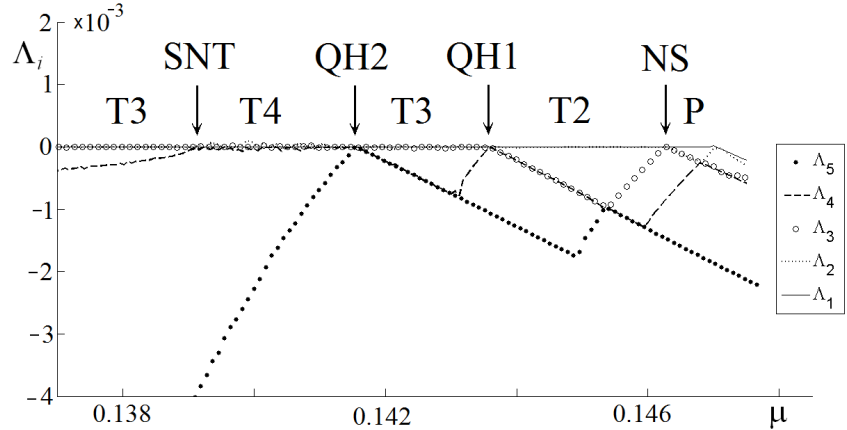}
\caption{Graph of five senior Lyapunov exponents vs. coupling parameter and
characteristic bifurcations points, $\Delta $=0.5.}
\end{figure}

Four-dimensional tori are also possible, however, they are less typical.
They are most reliably identified at high frequency detuning. Figure 3 shows
the corresponding graph for the five senior Lyapunov exponents at $\Delta
$=0.5. It allows to describe the bifurcations observed in the system. At
high coupling there is a stable equilibrium. Then from this state a stable
limit cycle is born via Andronov-Hopf bifurcation (AH), this cycle at the
Neimark-Sacker (NS) bifurcation point turns into a two-frequency torus. Now
the first $\Lambda _1 $ and the second $\Lambda _2 $ exponents
vanish\footnote{ The accuracy of the numerical calculations of zero Lyapunov
exponents for Figure 3 is about 10$^{ - 5 }-$10$^{ - 6}$, more accurate
calculations at several chosen points demonstrate that they are zero with
accuracy up to 10$^{ - 7}$-10$^{ - 8}$.}. Then at the QH1 bifurcation
point the three-frequency torus is born. Note that the type of bifurcation
in this case is determined by the behavior of the Lyapunov exponents. To the
right of the bifurcation point the third and fourth exponents coincide,
$\Lambda _3 = \Lambda _4 $, and at the bifurcation point they vanish. Then,
to the left, the third remains zero, while the fourth again becomes
negative. This behavior is typical for a quasi-periodic Hopf bifurcation
 responsible for the emergence of a three-frequency torus~\cite{21}. Then, as a
result of similar bifurcation QH2 the four-frequency torus is born. This
time, while approaching the point of bifurcation, the fourth and fifth
exponents coincide, $\Lambda _4 = \Lambda _5 $. With a further decrease of
coupling strength, however, there is a saddle-node homoclinic bifurcation of
tori SNT, and the dimension of the torus is reduced: there is
three-frequency resonant torus. In this case, the fifth Lyapunov exponent
remains negative all the time, indicating another (than QH) type of
bifurcation.

Figure 4 gives a portrait of four-frequency torus in the Poincare section
and the Fourier spectrum of the fourth oscillator in this case. It can be
seen that the spectrum consists of individual lines, wherein there are four
basic components. The remaining components correspond to the combination
frequencies.
\begin{figure}[h!]
\includegraphics[height=6cm, keepaspectratio]{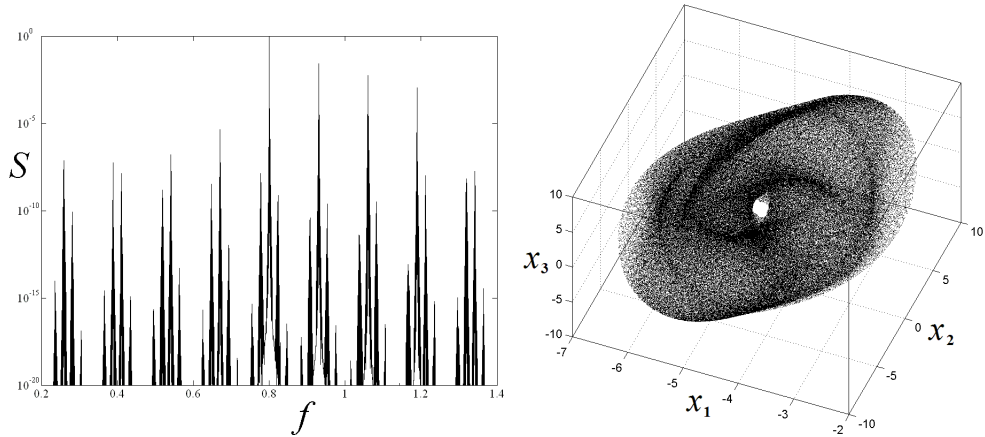}
\caption{Fourier spectrum and phase portrait of four-frequency torus in the
Poincare section, $\Delta $=0.5, $\mu $=0.14.}
\end{figure}

In the domain of lower frequency detuning $\Delta \approx 0.3 - 0.4$ the
Lyapunov chart in Figure 1 also renders four-frequency tori with four zero
Lyapunov exponents. However, a more detailed analysis using charting like
Figure 3 reveals also three-frequency tori with very small negative Lyapunov
exponent, so the caution is required to carefully distinguish two
situations.

It should be noted that our calculations show that the increase in the
number of chaotic oscillators to five leads to the possibility of
five-dimensional invariant tori. At the same time, however, it is advisable
to also introduce non-identity of the parameter $r$ responsible for the degree
of excitation of individual oscillators.

\section{Conclusion}
Thus, for a network of chaotic oscillators for large enough frequency
detuning it appears to be typical the presence of invariant tori of
different dimensions, including high-dimensional ones. These tori can
experience characteristic bifurcations, in particular, quasi-periodic Hopf
bifurcation and saddle-node bifurcation.

\section{Acknowledgement}
The authors thank Kuznetsov S.P. for useful discussions. The authors
acknowledge support of the RF President program for leading Russian research
schools NSh-1726.2014.2 and RFBR grant No 14-02-00085.

\textbf{\underbar{}}


\begin {thebibliography}{99}

\bibitem{1} G. Heinrich, M. Ludwig, J. Qian, B. Kubala, F. Marquardt, Phys. Rev. Lett. \textbf{107} (2011) ~043603.

\bibitem{2} M. Zhang, G.S. Wiederhecker, S. Manipatruni, A. Barnard, P. McEuen, M.
Lipson, Phys. Rev. Lett. \textbf{109} (2012) ~233906.

\bibitem{3} A. Temirbayev, Y.D. Nalibayev, Z.Z. Zhanabaev, V.I. Ponomarenko, M.
Rosenblum, Phys. Rev. E \textbf{87} (2013) ~062917.

\bibitem{4} E.A. Martens, S. Thutupalli, A. Fourri\`{e}re, O. Hallatschek, arXiv:1301.7608
(2013).

\bibitem{5} M.R. Tinsley, S. Nkomo, K. Showalter, Nature Phys. \textbf{8} (2012)~662.

\bibitem{6} V. Vlasov, A. Pikovsky, Phys. Rev. E \textbf{88} (2013)~022908.

\bibitem{7} T.E. Lee, M.C. Cross, Phys. Rev. Lett. \textbf{106} (2011)~143001.

\bibitem{8} T.E. Lee, H. R. Sadeghpour, Phys. Rev. Lett. \textbf{111} (2013)~234101.

\bibitem{9} C. Grebogi, E. Ott, J.A. Yorke, Physica D \textbf{15} (1985)~354.

\bibitem{10} P.S. Linsay, A.W. Cumming, Physica D \textbf{40} (1989)~196.

\bibitem{11} P.M. Battelino, C. Grebogi, E. Ott, J.A. Yorke, Physica D \textbf{39} (1989)~299.

\bibitem{12} C. Baesens, J. Guckenheimer, S. Kim, R.S. MacKay, Physica D \textbf{49} (1991)~387.

\bibitem{13} Yu.P. Emelianova, A.P. Kuznetsov, I.R. Sataev, L.V. Turukina, Physica D \textbf{244}
(2013)~36.

\bibitem{14} A.P. Kuznetsov, S.P. Kuznetsov, I.R. Sataev, L.V. Turukina, Phys. Lett.
A \textbf{377} (2013)~3291.

\bibitem{15} Yu.P. Emelianova, A.P. Kuznetsov, L.V. Turukina, I.R. Sataev, N.Yu.
Chernyshov, Commun. Nonlinear Sci. Numer. Simul. \textbf{19} (2014)~1203.

\bibitem{16} D. Paz\'{o}, E. S\'{a}nchez, M.A. Mat\'{\i}as, Int. J. Bifurcation Chaos
\textbf{11} (2001)~2683.

\bibitem{17} D. Paz\'{o}, M.A. Mat\'{\i}as, Europhys. Lett. \textbf{72} (2005)~176.

\bibitem{18} M.G. Rosenblum, A.S. Pikovsky, J. Kurths, Phys. Rev. Lett. \textbf{76} (1996)~1804.

\bibitem{19} G.V. Osipov, A.S. Pikovsky, M.G. Rosenblum, J Kurths., Phys. Rev. E \textbf{55}
(1997)~2353.

\bibitem{20} A. Pikovsky, M. Rosenblum, J. Kurths, Synchronization: A Universal Concept in Nonlinear Sciences, Cambridge University Press, 2001.

\bibitem{21} R.Vitolo, H. Broer, C. Sim\'{o}, Regul. Chaotic Dyn. \textbf{16} (2011)~154.

\end{thebibliography}
\end{document}